\documentclass[aps,twocolumn,floatfix,superscriptaddress]{revtex4-2}
\usepackage{graphicx}
\usepackage{dcolumn}
\usepackage{bm}
\usepackage{svg}
\usepackage{amssymb}
\usepackage{amsmath}
\usepackage{siunitx}
\usepackage{subfigure}
\usepackage{physics}
\usepackage{float}
\usepackage{multirow,tabularx}
\usepackage{sublabel}
\usepackage[utf8]{inputenc}
\usepackage{hyperref}
\usepackage{mathtools}
\usepackage{float}
\usepackage{csquotes}
\begin{document}


\title{Rashba-induced spin Hall response in a disordered \texorpdfstring{$WTe_2$}{WTe2} four-terminal structure}
\author{Swastik Sahoo}
\affiliation{Department of Electrical Engineering, Indian Institute of Technology Bombay, Powai, Mumbai-400076, India}




\author{Satadeep Bhattacharjee}
\affiliation{Indo-Korea Science and Technology Center, Bengaluru-560064, India}

\author{Bhaskaran Muralidharan}
\email[E-mail:~]{bm@ee.iitb.ac.in}
\affiliation{Department of Electrical Engineering, Indian Institute of Technology Bombay, Powai, Mumbai-400076, India}

\begin{abstract}
 The paramount acumen for controlling spin transport properties in nonmagnetic materials is the usage of spin-orbit coupling (SOC). We propose a model to calculate the spin hall angle (SHA) for the elemental transition metal dichalcogenide compound, $WTe_2$ entrenched on the intrinsic Rashba SOC. This model, is based on the Landauer-Buttiker formalism for quantum transport, and the $4$-terminal device setup with the presence of disorder from random onsite potential fluctuations. The SHA, including the mean and RMS values, also illustrate the mesoscopic oscillations, and the values obtained are $25\%$ and $30\%$, respectively. The variation pattern of charge and spin current, along with the mean and spin Hall conductance, can be a comparative measure for other TMDs and monolayer Xenes. To validate our outcomes, we compare our results with experimental data and numerically extract real-space simulation results based on the nearest-neighbor tight binding (NNTB) model. Also our results are in line with the scaling theory of localization. This work sets the stage to calculate the spin Hall angle and spin Hall conductivity for other elemental monolayer Xenes and TMDs, considering the intrinsic scattering mechanisms. 
An extension of this work will be to explore the possible spintronics applications for extrinsic scattering, including side-jump and skew-jump scattering processes.
 
\end{abstract}
\maketitle
\section{Introduction}
Two-dimensional (2D) materials, such as graphene and its analogues—silicene, germanene, and stanene exhibit exceptional electronic~\cite{castro2009electronic, novoselov2004electric} and mechanical properties~\cite{liu2007ab, lee2008measurement}, alongside notable advantages including scalability~\cite{levchenko2016scalable}, compatibility with already existing silicon fabrication technology~\cite{tao2015silicene}, and promising potential for applications in flexible electronics~\cite{sahoo2022silicene, sahoo2024density}. However, barely perceptible spin-orbit coupling (SOC)~\cite{kurpas2019spin} has stimulated the exploration of other (semi) conducting contenders such as hexagonal boron nitride (hBN)\cite{lee2012large} and transition-metal dichalcogenides (TMDs)\cite{wang2012electronics}. Monolayer TMDs are of percepts $MX_2$, where M is a $IVB-VIB$ transition metal atom (Ti and Zr, Nb and Ta, M and We) and X is a chalcogen (S, Se or Te). TMDs gain much interest due to showcasing their unique electrical, mechanical, optical, and thermal properties\cite{wilson1969transition,yoffe1973layer,yoffe1993low} and act as attractive material for spintronics applications due to their strong spin-orbit interaction~\cite{zollner2019strain, nan2019recent} and huge direct band gap~\cite{mak2010atomically}.   \\
\indent The TMDs are polymorphic with different crystal structures, including a trigonal prismatic 2H phase, a monoclinic 1T' phase, an orthorhombic Td phase, and a trigonal 1T phase\cite{jin2018phase}. Among TMDs, $WTe_2$ has been recognized as a rising member accommodating many unique characteristics such as low dimensionality, strong SOC, and structural mirage\cite{pan2018study}.  
Experimentally and theoretically, it has been established that $WTe_2$ has its minimum energy configuration in a distorted 1T structure\cite{augustin2000electronic}, following metallic-like transport properties. 
1T $WTe_2$ is a normal semimetal whereas 1T' $WTe_2$ is predicted to have nontrivial topological properties such as magnetoresistance\cite{zhou2022polymorph}, chiral anomaly\cite{xiong2015evidence} and  quantum spin Hall effect (QSHE)\cite{choe2016understanding,qian2014quantum}. \\
\indent Quantum spin Hall effect is a popular phenomenon in TMDs and several theoretical and experimental observations have been performed on $WTe_2$ to realize its potential in spintronics and its fundamental applications like spin Hall angle (SHA). 
These observations show a broader range of SHA and the scrutiny is on the hybrid structures of $WTe_2$. So, given the strong SOC and the potential SHE, we thought of exploring SHA in a pristine monolayer $WTe_2$ with some disorder, which highlights the novelty and rarity of the paper. Moreover, motivated by references \cite{da2022spin,ganguly2018magnetic}, we analyze the behavior of SHA in monolayer $WTe_2$ in the presence of uniform (variable) SOC (energy) and the other way around. 
Finally, to validate our outcomes, we compare our results with experimental data and numerically extract real-space simulation results based on the nearest-neighbor tight-binding (NNTB) model. This work will serve to be a template to calculate SHA for the TMDs, considering the intrinsic scattering mechanism and  further can be extended to numerous other material combinations and extrinsic scattering mechanisms.  \\  
\indent The governance of information through spin was discovered by Datta and Das\cite{datta1990electronic}, and the objective was to use spin as a message carrier in place of charge, that made high-speed computing possible. 
Elaborately, when a longitudinal charge current flows through a region of strong SOC, it generates a transverse spin current\cite{schliemann2006spin,sinova2015spin} and this conversion 
is a measurable quantity, called SHA ~\cite{wang2015perpendicular,seifert2018terahertz}. As SHE goes hand in hand with our area of research, henceforth, we have considered the monoclinic 1T' phase having the advantage of most stable structure at room temperature $(T > 294K)$. \\
\indent The remainder of this article is organized as follows. Section II briefly focuses on the computational method of SHA fluctuations, established on the Landauer-Buttiker (LB) model. In section III, we analyze the obtained charge and spin current plots, conductance plots along SHA plots in presence of disorders in detail. The numerical calculations are performed using KWANT software\cite{groth2014kwant}. Finally, our conclusions are presented in section IV. 

\section{Structure and Design} 
\begin{figure*}[t]
     \centering
\includegraphics[width=1.00\textwidth]{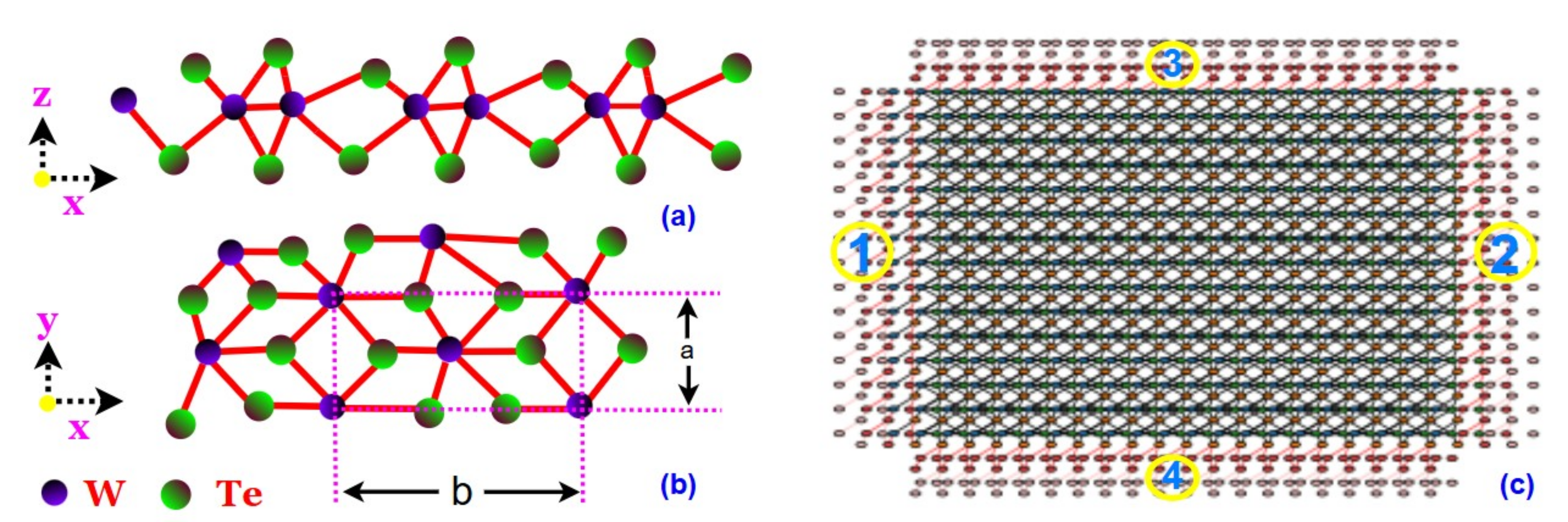}
     \caption{(a) Side view, (b) Top view  of crystal lattice structure of monolayer 1T' $WTe_2$. \lq{a}\rq{}  and \lq{b}\rq{} represents the lattice constants. Its rectangular unit cell is indicated by dashed purple line.  (c) Spin Hall device design of single layer $WTe_2$. Scattering sample with disorder is connected to four semi-infinite leads numbered from 1 to 4. Lead number 1 and 2 represent longitudinal leads and 3 and 4 are transverse leads. }
    \label{fig:P03_1}
 \end{figure*}
In this section, we briefly wade into the structure and device fundamentals of monolayer TMD. $WTe_{2}$ consists of W (transition metal) and Te (chalcogen), where each tungsten layer is sandwiched between two tellurium layers and forms strong ionic bonds\cite{chang2016prediction}. In the area of our interest we consider that, $WTe_2$ exists in two different phases. The 1T phase has a regular hexagonal structure, whereas the 1T' phase comes from the lattice distortion of the 1T phase\cite{tang2017quantum}.\\ 
\indent This model uses a rectangular lattice to approximate the monolayer tungsten ditelluride $(WTe_2)$. The side view and top view of the crystal structure of monolayer $1T'$ phase $WTe_2$ is shown in fig~\ref{fig:P03_1}(a) and fig~\ref{fig:P03_1}(b) respectively and this $1T'$ semi-metallic phase exhibits a distorted octahedral structure\cite{chang2016prediction}. The lattice constants are $a=3.496$ {\AA} and $b=6.282$ {\AA} \cite{dawson1987electronic}, representing the periodicity in the xxx- and yyy- directions respectively. $WTe_2$ is a layered material with a strong in-plane anisotropy\cite{song2016plane}. Since the transport calculation majorly focuses on monolayer $WTe_2$, we primarily need the in-plane rectangular Bravais lattice parameters in x- and y- directions. The third vector is only relevant in bulk $WTe_2$ when stacking layers and no interlayer coupling in monolayer $WTe_2$\cite{kim2015anomalous} are there. The model includes three sublattices: tungsten $(W)$, tellurium 1 $(T_{e}1)$, and tellurium 2 $(T_{e}2)$. The bond lengths are specified as $2.74$ {\AA} and $2.84$ {\AA}, between $W-T{e}1$ and $W-T{e}2$ respectively\cite{lee2015tungsten}. The difference in bond lengths suggests an asymmetry in W-Te bonding, contributing to the material's unique electronic properties such as strong SOC and topologically nontrivial states\cite{keum2015bandgap}. Fractional shifts are applied to match these bond lengths accurately. Further, the interatomic hopping between W and Te atoms includes nearest-neighbor hopping along both x- and y-directions, forming the basis of the tight-binding (TB) model, and their values are listed as 0.99 eV and 0.03 eV\cite{hu2021realistic}. These values suggest that electron movement along the x-direction is significantly stronger than in the y-direction, justifying the fact that, $WTe_2$ is a highly anisotropic material with quasi-$1D$ electronic behavior. \\
\indent Fig.\ref{fig:P03_1}c represents a spin Hall device with four semi-infinite leads connected to a scattering region with the disorder. Vertical and horizontal leads are attached to the system for transport calculations. The left and right leads represent the source and the drain, respectively. The top lead is for the spin accumulation detection, and the bottom is for the spin current direction. These leads are constructed using translational symmetry in the xxx- or yyy directions. These leads enforce open boundary conditions and connect the system to reservoirs, enabling the computation of charge and spin currents. The charge current is calculated using the transmission matrix between the source and drain, and that for the spin current is from the transmission matrix between the top and bottom leads, reflecting spin-polarized transport. \\
\indent This current calculation is followed by the calculation of SHA. Simulations for our TB model are conducted over a range of disorder strength and energy values, with multiple samples for averaging. Previously, reference \cite{ramos2018spin} developed an analytical expression of mean spin Hall current for graphene device at charge neutrality point (CNP), based on random matrix theory (RMT)\cite{beenakker1997random}. Moreover, da silva \textit{et al}\cite{da2022spin} showed the calculation of SHA by implementing central limit theorem to develop the ensemble averages of the parameters. We have adopted numerical mean method based on our TB model, as former one is smoother and reduces noise but can not capture disorder fluctuations effectively and the later one will capture that directly in SHA values. In addition to that, TB model is more realistic for specific material like $WTe_2$ whereas RMT provides universal statistical predictions for transport properties in mesoscopic physics. Overall, our model represents a purely electronic transport systems and any thermal effects and electron-phonon interactions are ignored for simplicity. 
\section{Computational details}
We designed a spin Hall device with four semi-infinite leads connected to a scattering region with disorder. We propose a model to calculate the SHA for layered Weyl semi-metal like $WTe_2$. This model is based on LB formalism, which is a widely used approach for quantum transport calculations in mesoscopic systems\cite{nikolic2005mesoscopic,bardarson2007mesoscopic}. This method treats electron transport ballistically and expresses current in terms of transmission probabilities. For a multi-terminal system like this, the current at a particular lead is given by: 
\begin{equation} \label{P01_eq14}
I_{\alpha}= \frac{e^2}{h}\sum_{\beta=1}^{4} (T_{\alpha\beta}V_{\alpha}-T_{\beta\alpha}V_{\beta})
\end{equation}
here, $T_{\beta\alpha}$ is the transmission coefficient from lead $\alpha$ to $\beta$. $V_\alpha$ and $V_\beta$ are the applied voltages at the respective leads and $I_{\alpha}$ is the total current computed by summing over all connected leads. The potential difference between two horizontal leads, generates a pure longitudinal charge current. The spin-up carriers are condensed to one side of the scattering region and the spin-down carriers are condensed to another side, resulting in transverse spin current. The transmission coefficient between two leads is extracted as: 
\begin{equation}
T_{\beta\alpha}= \sum_{n,m} \mid S_{\beta n,\alpha m}\mid ^2, ~~~S=\begin{bmatrix}
r_{11} & t_{12} & t_{13} & t_{14}\\
t_{21} & r_{22} & t_{23} & t_{24} \\
t_{31} & t_{32} & r_{33} & t_{34}\\
t_{41} & t_{42} & t_{43} & r_{44}\\
\end{bmatrix}
\end{equation}
where n, m are the quantum states in the leads. S represents the scattering matrix 
 and a generic expression is presented here. The elements in the matrix conveys the quantum state of the leads. Each element in $T_{mn}$ represents the conductance contribution from lead m to lead n. Ideally, net charge current vanishes in the transverse direction due to the absence of an external charge bias across the corresponding leads\cite{nikolic2007extrinsically,bardarson2007mesoscopic} but in realistic system with disorder, there might be small non-zero charge current. Spin current is obtained from spin-dependent transmission operator between different leads correspond to spin-filtered transport in transverse leads. The mathematical expression is: 
 \begin{equation} \label{P01_eq14}
I_{spin,i}= \frac{e^2}{h}\sum_{j\neq1} Tr [\sigma_s(t_{ji}^\dag\ t_{ji})]
\end{equation} 
where, $\sigma_s$ is the Pauli matrices with polarization direction as x, y, z and trace operation extracts spin-resolved transmission. \\
\indent As mentioned in the previous section and as per the TB model, we express the generic equations of SHA fluctuations around the mean.
\begin{equation}
 \langle \theta_{SHA}  \rangle = \frac{1}{N} \sum_{i=1}^{N} \theta_{SHA,i} 
\end{equation}

\begin{equation}
 \theta_{SHA}^{RMS} = \sqrt{\frac{1}{N} \sum_{i=1}^{N} \left[ \frac{I_{spin,i}}{I_{charge,i}}-\langle \theta_{SHA}  \rangle \right] ^2}
\end{equation}
here, N represents number of disorder realizations and rest of the parameters have their usual meanings.

\indent To calculate the SHA, we have performed numerically real-space simulation.The above $WTe_2$ model visualizes the lattice, showing the atomic positions and hopping terms. This $WTe_2$ model can be used to study: (i) spintronics properties, including spin Hall and inverse spin Hall effects, (ii) Disorder effects on electronic transport, (iii) Relationship between SOC strength, disorder and spin-related transport phenomena. The above implementation highlights the interplay between SOC, disorder and the unique lattice structure of $WTe_2$, making it a powerful tool for the theoretical and computational investigations into spintronics materials.  

\section{Results and discussion}
For our TB model based $WTe_2$, we performed numerically accurate real-space simulation for SHA. For that, we have defined the Hamiltonian, which has included 3 components and it is expressed as: $H=H_{h}+H_{R}+H_{d}$. Here, $H_h$ is the interatomic hopping between W and Te atoms. This includes the nearest-neighbor hopping along both x-and y-directions, forming the basis for the TB model, and the hopping parameters are direction dependent due to the anisotropic nature of $WTe_2$. $H_R$ represents the spin-orbit coupling Hamiltonian, that is included in the hopping function of $WTe_2$ and the SOC term is added via the Pauli Matrices, giving rise to spin-dependent effects. The SOC term, effectively couples the spins and is essential for simulating spin Hall effects, which are a prominent feature of TMDs like $WTe_2$. $H_d$ represents the onsite disorder using an Anderson-type disorder potential. It introduces random variation in the potential at each site of the $WTe_2$ lattice. The above mentioned Hamiltonian are mathematically deciphered as follows: 
\begin{equation}
H_h = \sum_{i,j} t_{i,j} c_i^{\dagger} c_j   
\end{equation}
\begin{equation}
H_R = i\lambda_{SOC} (\sigma_z  \otimes \sigma_y)
\end{equation}
\begin{equation}
H_d = U_i c_i^\dagger c_i
\end{equation}
where, indices $i$ and $j$ run over all lattice sties. $c_i^{\dagger}$ and $c_j$ are the creation and annihilation operators at sites $i$, $j$. $t_{i,j}$ is the hopping energy between sites i and j. $\lambda_{SOC}$ is the Rashba copuling SOC strength between sites $i$ and $j$. The tensor product between Pauli matrices introduces SOC effects. $U_i$ is a random on-site potential drawn from a uniform distribution (-U/2, U/2). $c_i^\dagger c_i$ represents electron density at site $i$. The electro-chemical potential difference between  source and drain, Rashba SOC and disorder strength are given in the units of t, while the length and width of the device W=40, L=30 are given in the units of $WTe_2$ lattice constants. All the numerical results were computed over 500 disorder realizations. The numerical calculations were implemented in KWANT software.\\
\indent Furthermore, in case of a four terminal device, the longitudinal and spin Hall conductance are defined as follows\cite{ganguly2018magnetic}: 
\begin{equation}
    G= \frac{I_2}{V_1-V_2}, ~~~  
    G_{SH}^\alpha = \frac{\hbar}{2q}\left[\frac{I_3^\alpha}{V_1-V_2}\right]
\end{equation}
Here, $I_2$ is the charge current flowing through terminal 2 and $I_3^\alpha$ is the spin current polarized in a particular direction $\alpha$ (where $\alpha$= x, y and z, each polarization direction is inspired from the respective Pauli Matrices) and flowing through terminal 3. The longitudinal conductance and the spin Hall conductance are measured in the units of $\frac{e^2}{\hbar}$ and $\frac{e}{4\pi}$ respectively. The voltage applied at respective leads is denoted likewise. As we know, conductance will follow a similar trend like current, so the variation of longitudinal conductance is shown in supplementary information and our prime focus will be on the variation pattern of spin Hall conductance along the three axes.   
\\
\indent We have demonstrated the simulated numerical calculations for two different configurations. The first configuration is for $WTe_2$ spin Hall device with uniform Rashba SOC, where we keep the SOC strength uniform for the whole lattice while varying the disorder strength and energy. For the second case, we vary the Rashba SOC and disorder strength while keeping a uniform energy throughout the lattice structure. \\
\subsection{Uniform Rashba SOC}
\begin{figure*}[t]
     \centering
\includegraphics[width=1.05\textwidth]{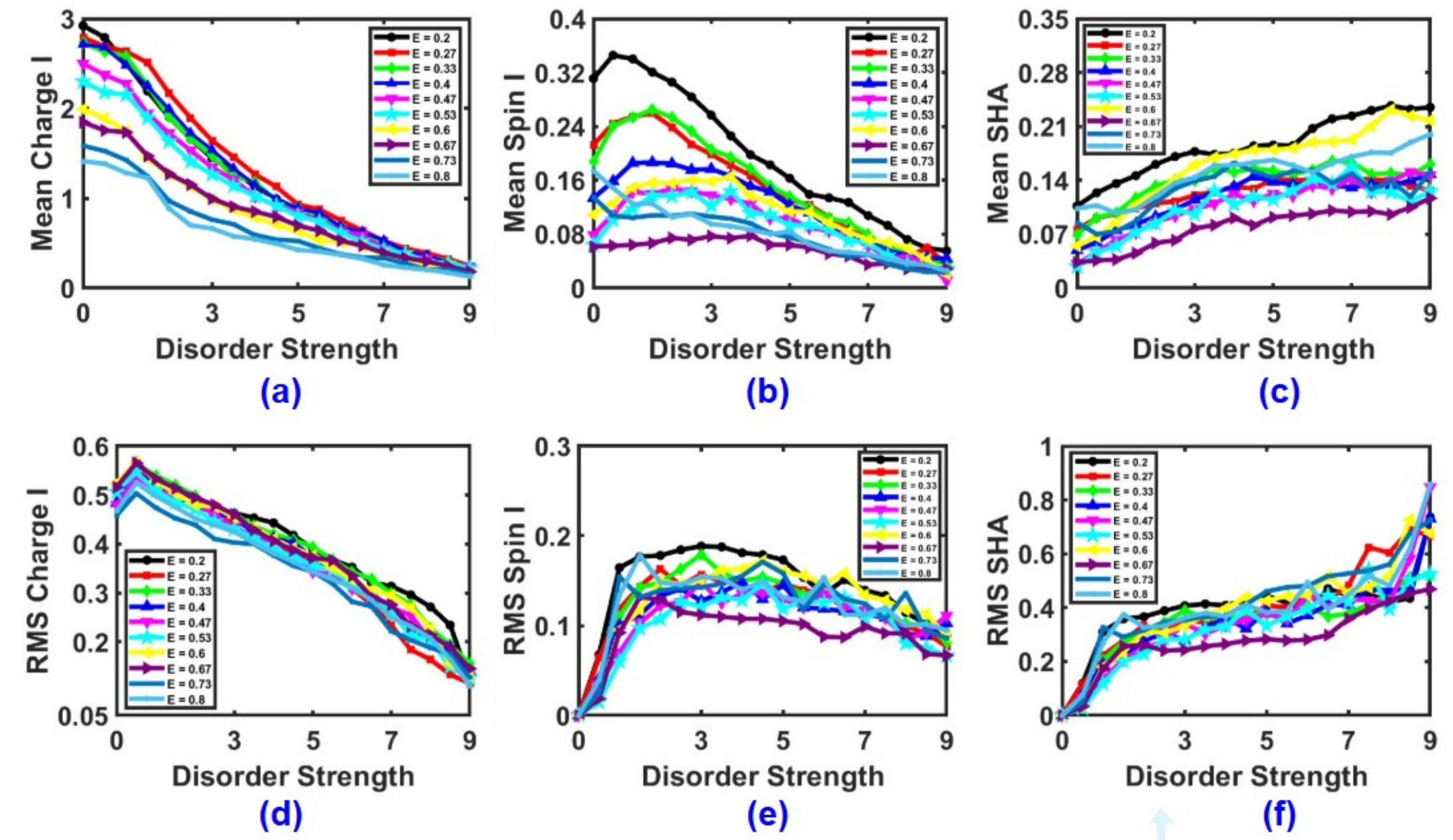}
     \caption{Plots of Mean (a) Charge current, (b) Spin  current, (c) Mean spin Hall angle, RMS (d) charge current (e) spin current, (f) RMS spin Hall angle as a function of disorder strength for different values of energy (0.2 eV - 0.8 eV) at a fixed SOC value of 50 meV.}
    \label{fig:P03_fig2}
 \end{figure*}

In this section, we inspect the spin Hall parameters and the variation of conductance with the uniform Rashba SOC to get a glimpse of transport properties in the nanoscale regime. The SOC strength is fixed at $50$ meV for both current and conductance, and the disorder is introduced by random on-site potential with uniform distribution.  \\
\indent Figure~\ref{fig:P03_fig2} shows the variation of mean (a) charge, (b) spin current, rms (d) charge, (e) spin current and then (c) mean, and (f) RMS spin Hall angle with respect to disorder strength for varying energy from 0.2 to 0.8 eV keeping an uniform Rashba SOC as 50 meV. The disorder strength is introduced by an electrostatic potential. This low energy value corresponds to the proximity of the Fermi level, band structure efficiency and computational efficiency. Similarly, the reason to choose this particular value of Rashba SOC corresponds to the balance between computational efficiency and physical efficiency. Moreover, this value is good enough for observing spintronic phenomena.\\
 \indent The mean charge current is high for low disorder strength and as the disorder increases, the charge current decreases due to enhanced scattering. This decrement is not necessarily monotonic. The system moves towards Anderson localization regime and the current saturates as the disorder strength reaches its peak value. Spin current is primarily due to intrinsic SOC effects. The mean spin current, at the beginning, displays a tiny increment at initial level of  disorder levels, followed by a peak, and then constant decrement at higher disorder strength. After the current attains the maximum value, the disorder-induced scattering becomes dominant, resulting in reducing the spin current. Mean SHA as usual, represents the efficiency of charge-to-spin current conversion, it is mainly determined by intrinsic SOC and the trend is non-monotonic. At low energy value, SHA starts high and remains relatively elevated. At higher energies, SHA increases slowly and follows a steady increment. This draws the conclusion that, at increased disorder, the spin-charge conversion efficiency is enhanced. \\
 \indent The RMS charge current plot can be explained in three transitions. At low disorder, RMS charge current is small as the transportation is nearly ballistic and fluctuations are minimal.Afterwards, this current attains its peak value due to strong fluctuations caused by multiple scattering paths. At high disorder, the current decreases, leading to a nearly static transmission. The behavior of RMS spin current can also be substantiated using 3 steps, similar to that of RMS charge current. Initially, the current is at its lowest value due to clean transport. Then the spin current increases significantly, indicating strong spin-dependent scattering and the values remain finite at higher disorders. For RMS SHA, fluctuation is lowest initially and it will peak at intermediate disorder;  gradually the fluctuation is high at strong disorder due to dominant Rashba SOC. These large fluctuations imply the spin-to-charge conversion efficiency varies significantly with scattering events. Higher energies, which initially had subsequently lower SHA values, exhibit increased fluctuations at higher disorder strengths, reflecting the greater susceptibility to scattering variations.\\
 \begin{figure*}[t]
     \centering
\includegraphics[width=1.05\textwidth]{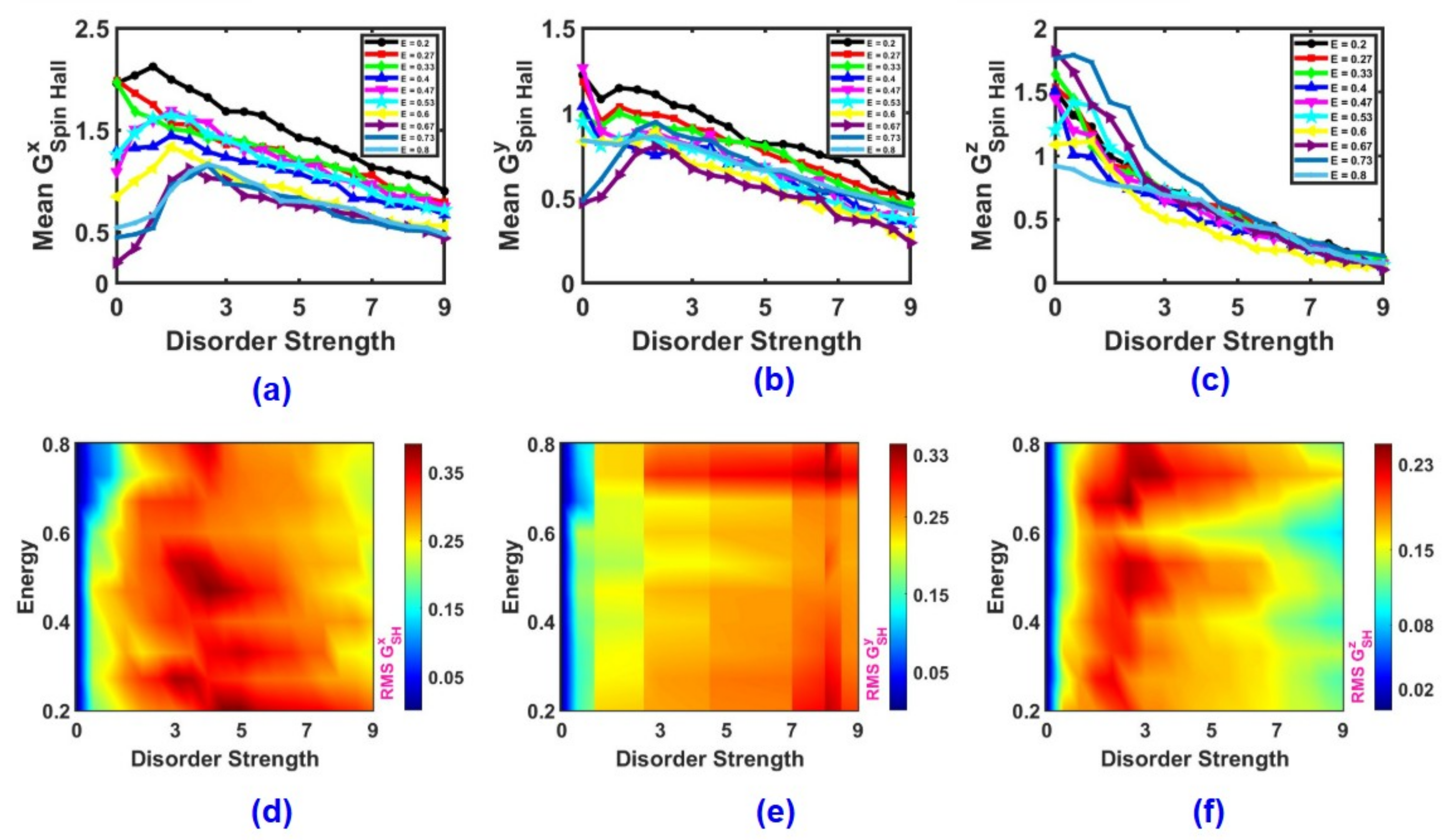}
     \caption{Plots of Mean spin Hall conductance along (a) x- (b) y- and (c) z- direction and plots of RMS spin Hall conductance along (d) x, (e) y and (f) z-direction as a function of disorder strength for different values of energy (0.2 eV - 0.8 eV) at a fixed SOC value of 50 meV.}
    \label{fig:P03_fig4}
 \end{figure*}
 \indent The behavior of all three components of spin Hall conductance in the presence of disorder, with variable energy is shown in Fig.~\ref{fig:P03_fig4}. We have considered both mean and RMS values to get a fair idea about the variation pattern as to how robustly this spin-charge conversion occurs as disorder and energy alter. Fig.~\ref{fig:P03_fig4}(a) represents the mean spin Hall conductance along the x-direction. Lower energy plots yield higher mean values, indicating sensitivity to intrinsic spin-orbit textures and higher energy depicts reduced spin conversions. The plots display enhancement at moderate disorder and eventually decline at higher disorder, indicating localization effects. So precisely, mean $G_{SH}^x$ may be less sensitive to intrinsic Rashba effects, primarily due to disorder-induced symmetry breaking. Fig.~\ref{fig:P03_fig4}(b) represents the mean spin Hall conductance along y-direction. Energy-dependent conductance plots are moderately varied. With increasing disorder, a kind of plateau is followed by a subsequent decline at higher disorders, but less pronounced compared to that along the x-direction.  Fig.~\ref{fig:P03_fig4}(c) represents the mean spin Hall conductance along the z-direction. This plot depicts the highest amount of variation compared to other axes. Higher energy conductance plots show reduced but stable behavior. The higher disorder also signifies a more significant drop. So, this dominance is intrinsic to Rashba interaction, strongly influenced by the intrinsic SOC. \\
 \indent RMS spin Hall conductance plots will provide insights into the robustness or sensitivity of the spin Hall conductance under the effect of disorder and energy variation. For the RMS part, we have drawn the heat map to exhibit the fluctuation clearly as in general, they will have a larger range. Fig.~\ref{fig:P03_fig4}(d) represents the RMS spin Hall conductance along the x-direction. From the heat map, we can easily infer that, the fluctuations attain their maximum value at moderate disorder, minimum value at low disorder, and moderate value at higher disorder strength. Fig.~\ref{fig:P03_fig4}(e) represents the RMS spin Hall conductance along the y-direction. The fluctuations go alongside higher energy and higher disorder strength. Maximum value is obtained at higher energies, even at moderate disorder. This plot infers relatively low sensitivity at higher disorder compared to the x-direction, rendering more robust spin transport along the y-direction. Fig.~\ref{fig:P03_fig4}(f) represents the RMS spin Hall conductance along the z-direction. This plot displays the lowest fluctuation at low and higher disorder strength, and at moderate disorder, fluctuations significantly increase but the magnitude is least among all. So, this detailed analysis underscores how fine-tuning disorder strength and energy can optimize the performance of $WTe_2$ based devices leveraging intrinsic Rashba physics which can be extended to extrinsic disorder-mediated mechanisms. 

 \subsection{Malleable Rashba SOC}
 \begin{figure*}[t]
     \centering
\includegraphics[width=1.05\textwidth]{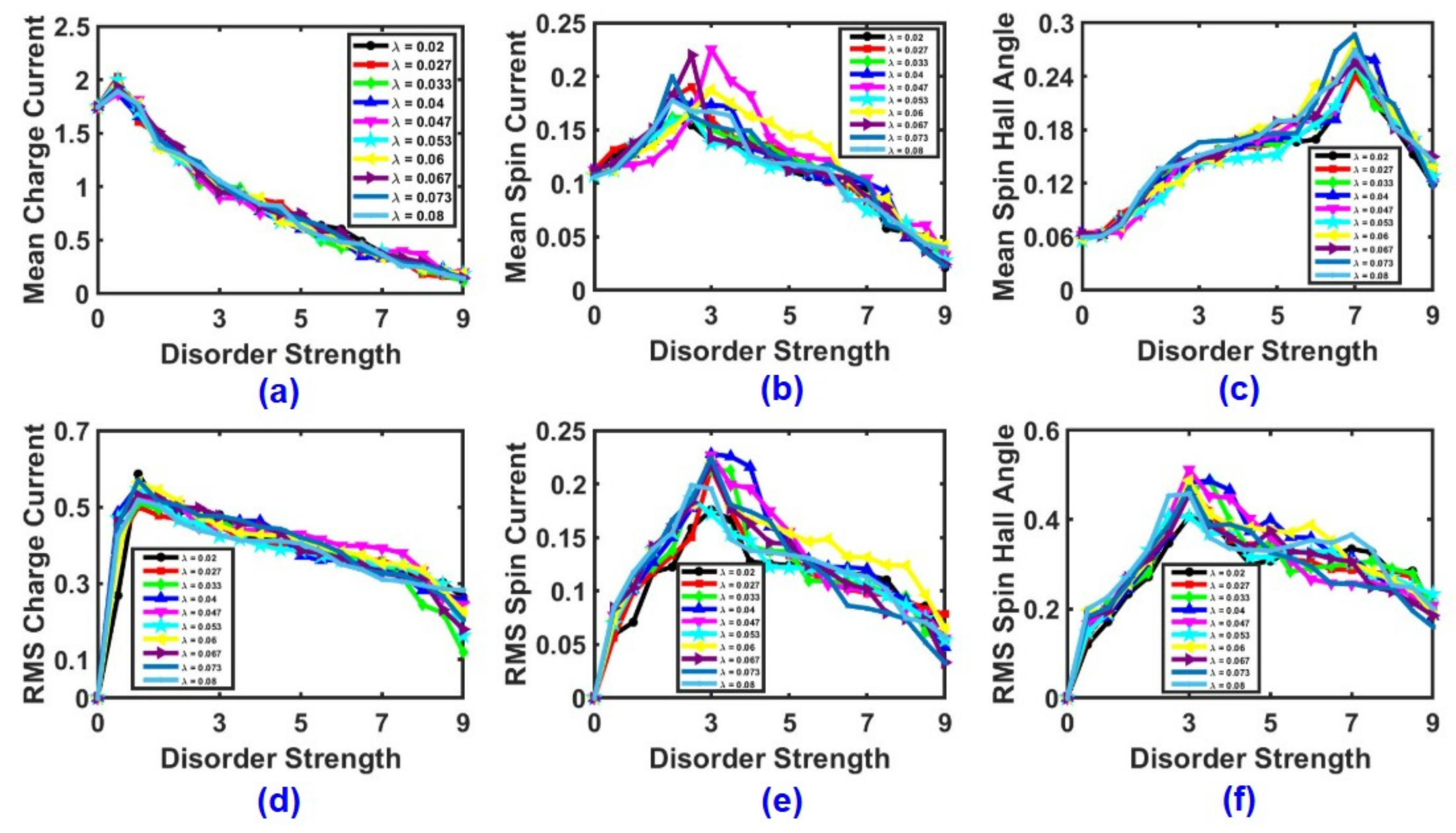}
     \caption{Plots of Mean (a) Charge current, (b) Spin  current, (c) Mean spin Hall angle, RMS (d) charge current (e) spin current, (f) RMS spin Hall angle as a function of disorder strength for different values of Rashba SOC (0.02 eV - 0.08 eV) at a fixed energy value of 0.6 eV.}
    \label{fig:P03_fig3}
 \end{figure*}
\begin{figure*}[t]
     \centering
\includegraphics[width=1.05\textwidth]{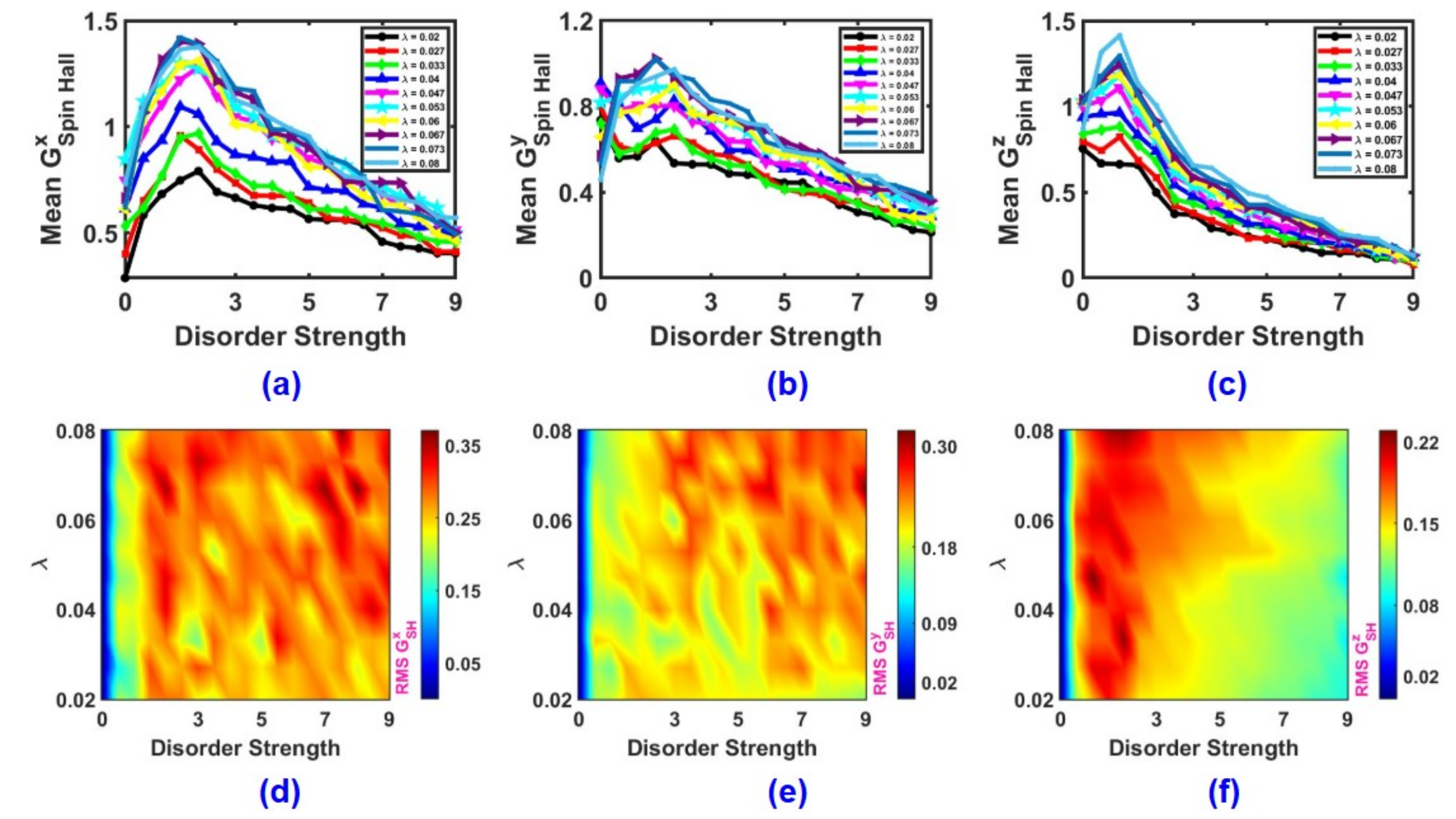}
     \caption{Plots of Mean spin Hall conductance along (a) x (b) y and (c) z- direction and plots of RMS spin Hall conductance along (d) x, (e) y and (f) z-direction as a function of disorder strength for different values of SOC (0.02 eV - 0.08 eV) at a fixed energy of 0.6 eV.}
    \label{fig:P03_fig5}
 \end{figure*}

In this section, we inspect the spin Hall parameter and conductance variation for fixed energy with variable Rashba SOC. Figure ~\ref{fig:P03_fig3} shows the variation of mean (a) charge, (b) spin current, RMS (d) charge, (e) spin current and then
(c) mean, and (f) RMS spin Hall angle with respect to disorder strength for varying Rashba SOC from 20 to 80 meV, 
keeping a uniform energy value of 0.6 eV. The disorder strength is introduced by an electrostatic potential. These plots provide insights into how varying Rashba SOC and disorder strength affect charge current, spin current, and spin Hall angle. Fig.~\ref{fig:P03_fig3}(a) represents the mean charge current, which decreases monotonically as disorder increases. Initially, at low disorder, the charge current is relatively high, signifying efficient charge transport. Then as the disorder increases, the current reduces steadily due to enhanced scattering and reduced mobility. We can infer an important point that, variations in Rashba SOC have negligible effects on mean charge current, indicating that Rashba effects primarily influence spin rather than charge transport directly. Fig.~\ref{fig:P03_fig3}(b) represents the mean spin current, depicting non-monotonic behavior with increasing disorder. The pattern is initially low at small disorder strength. Then it peaks notably around the moderate disorder, reaching maximum efficiency, and then declines at higher disorder. Increased Rashba SOC slightly enhances the magnitude of the spin current peak, but the excessive disorder leads to suppressed spin transport. Fig.~\ref{fig:P03_fig3}(c) represents the mean SHA, inferring clear non-monotonic disorder, similar to the mean spin current. This plot depicts that a higher Rashba SOC slightly increases SHA, reinforcing efficient conversion but remains broadly similar across Rashba variations, concluding robustness against small Rashba changes at fixed energy. SHA is initially modest, and at intermediate disorder, the plot suggests a strong enhancement of spin-to-charge conversion efficiency. \\
\indent Coming to the lower row, Fig.~\ref{fig:P03_fig3}(d) represents the RMS charge current. RMS fluctuations initially increase rapidly from zero disorder, reaching a peak value. Afterward, RMS fluctuations steadily decrease as the disorder increases further, inferring uniform localization significantly reduces the fluctuations. Again, variations across Rashba SOC are minor, confirming a minimal influence of Rashba on charge current fluctuations. Fig.~\ref{fig:P03_fig3}(e) represents the RMS spin current fluctuations, pronounced peak at intermediate disorder, mirroring the mean spin current trend. After attending the maximum value, the RMS value decreases gradually with increasing disorder due to localization effects affecting the spin transport. Higher Rashba values slightly amplify these fluctuations at peak points, indicating greater sensitivity of spin transport to SOC magnitude at moderate disorder. Fig.~\ref{fig:P03_fig3}(f) represents the RMS SHA fluctuations. These exhibit strong peaks around moderate disorder strength. Initially, fluctuations are low, rising significantly at intermediate disorder. Beyond peak values, SHA values reduce along the higher disorder strength. Further, Rashba variations enhance the peak magnitude, indicating stronger SOC dependence for SHA fluctuations. Overall, this comprehensive analysis indicates the critical role of disorder and Rashba SOC in tuning spin transport and spin-charge conversion properties, which is crucial for developing robust TMDC-based devices. \\
\indent We now analyze the variation pattern of spin Hall conductance along x, y, and z-direction for varying disorder strength with varying Rashba SOC having a constant energy value in Fig.~\ref{fig:P03_fig5}. 
Fig.~\ref{fig:P03_fig5}(a) represents the mean spin Hall conductance along the x-direction, showing a non-monotonic dependence on disorder strength. At low disorder, mean conductance rises rapidly and attends peak values. Further, an increase in disorder strength, results in a reduced conductance value significantly, indicating a detrimental effect of high disorder on spin transport. Enhanced Rashba SOC generally leads to increased peak values, demonstrating that stronger SOC can enhance spin-charge conversion efficiency along the x-axis. Fig.~\ref{fig:P03_fig5}(b) represents the mean spin Hall conductance along y-direction. Here, similar non-monotonic trends are observed, but peak heights are relatively lower compared to those along the x-direction. Peak conductance values are attained at moderate disorder and then the values gradually reduce as the disorder increases further. Rashba SOC again moderately enhances the peak magnitude, highlighting the role of SOC in spin Hall effects along the y-axis. Fig.~\ref{fig:P03_fig5}(C) represents the mean spin Hall conductance along the z-direction. This plot exhibits monotonic behavior, which is distinct from the previous two directions. The values are high initially at low disorder strength and further, they decrease steadily. Higher Rashba SOC does not alter the overall declining trend significantly. \\
\indent The RMS fluctuations of spin Hall conductance are illustrated by heat maps, providing insights into the sensitivity and stability of spin Hall effects. Fig.~\ref{fig:P03_fig5}(d) represents the RMS spin Hall conductance along the x-direction. Strong fluctuations appear at moderate disorder strength and low fluctuations occur at very low or very high disorder strengths, indicating stable spin Hall effects under extreme conditions (either clean or highly disordered regimes). Higher Rashba SOC enhances the magnitude of RMS fluctuations at moderate disorder. 
Fig.~\ref{fig:P03_fig5}(e) represents the RMS spin Hall conductance along the y-direction. This plot exhibits a similar pattern to that of x-direction, though fluctuations are relatively smaller. Peak values occur at moderate disorder and RMS fluctuation magnitude increases slightly with larger Rashba SOC, indicating SOC-dependent instability in the y-direction spin transport. Fig.~\ref{fig:P03_fig5}(f) represents the RMS spin Hall conductance along the z-direction. This plot exhibits relatively lower RMS fluctuation compared to that along x- and y-directions. Fluctuations peak at low to moderate disorder and diminish quickly with increasing disorder. Rashba SOC has less influence on fluctuations along the z-direction, highlighting reduced sensitivity and better stability of spin transport in this direction under varied conditions.
\setlength{\tabcolsep}{7pt}
\begin{table*}[!t]
\centering
\begin{tabular}{c c c}\\ [0.08 cm]
 \hline\hline
$WTe_2$ crystal structure & Value of SHA   & Reference  \\
\hline
$WTe_2$ in $T_d$ phase at room temperature & $17\%$ & \cite{zhao2020observation}\\ [0.03 cm]
Monolayer $1T'$ $WTe_2$ (Varying energy and constant SOC) & $25\%$ & This Work \\[0.03cm]
Monolayer $1T'$ $WTe_2$ (Varying SOC and constant energy) & $30\%$ & This Work \\[0.03cm]
Orthorhombic crystal structure of $T_{d}-WTe_2$ & $54\%$ & \cite{zhou2019intrinsic}\\[0.08 cm]
\hline \hline  
    \end{tabular}
    \caption{Comparison of spin Hall angle for various configuration of $WTe_2$ crystal structure}
    \label{tab:my_label}
\end{table*}
\subsection{Implication of the results}
Table~\ref{tab:my_label} summarizes the findings of the research. We obtain a maximum mean SHA of $25\%$ for varying energy and constant SOC, and a maximum of $30\%$ for varying Rashba SOC and constant energy. These results are well within the range as mentioned in table~\ref{tab:my_label}. However, for  $WTe_2$-graphene like heterostructure, maximum SHA from the experimental findings is $1.4\%$ \cite{zhao2020observation}, which defines the broader range of SHA for $WTe_2$ under various conditions. RMS SHA fluctuations go up to $80\%$ for varied energy whereas the same goes up to $50\%$ for varied Rashba SOC. The reason is that energy variations can significantly alter spin relaxation lengths and spin-mixing rates, causing a pronounced fluctuation in the spin polarization and hence more fluctuation in SHA. Again, changing Rashba SOC predominantly influences spin precession angles rather than drastically modifying spin relaxation regimes, causing milder fluctuations. So, the former system is very sensitive to disorders and small perturbations can cause significant variations in the efficiency of spin current generation. On the contrary, the latter system is comparatively robust and stable against variations and disorders. \\
\begin{figure}[!htbp]
 \begin{minipage}[!htbp]{0.47\textwidth}
     \centering
\includegraphics[width=1.05\textwidth]{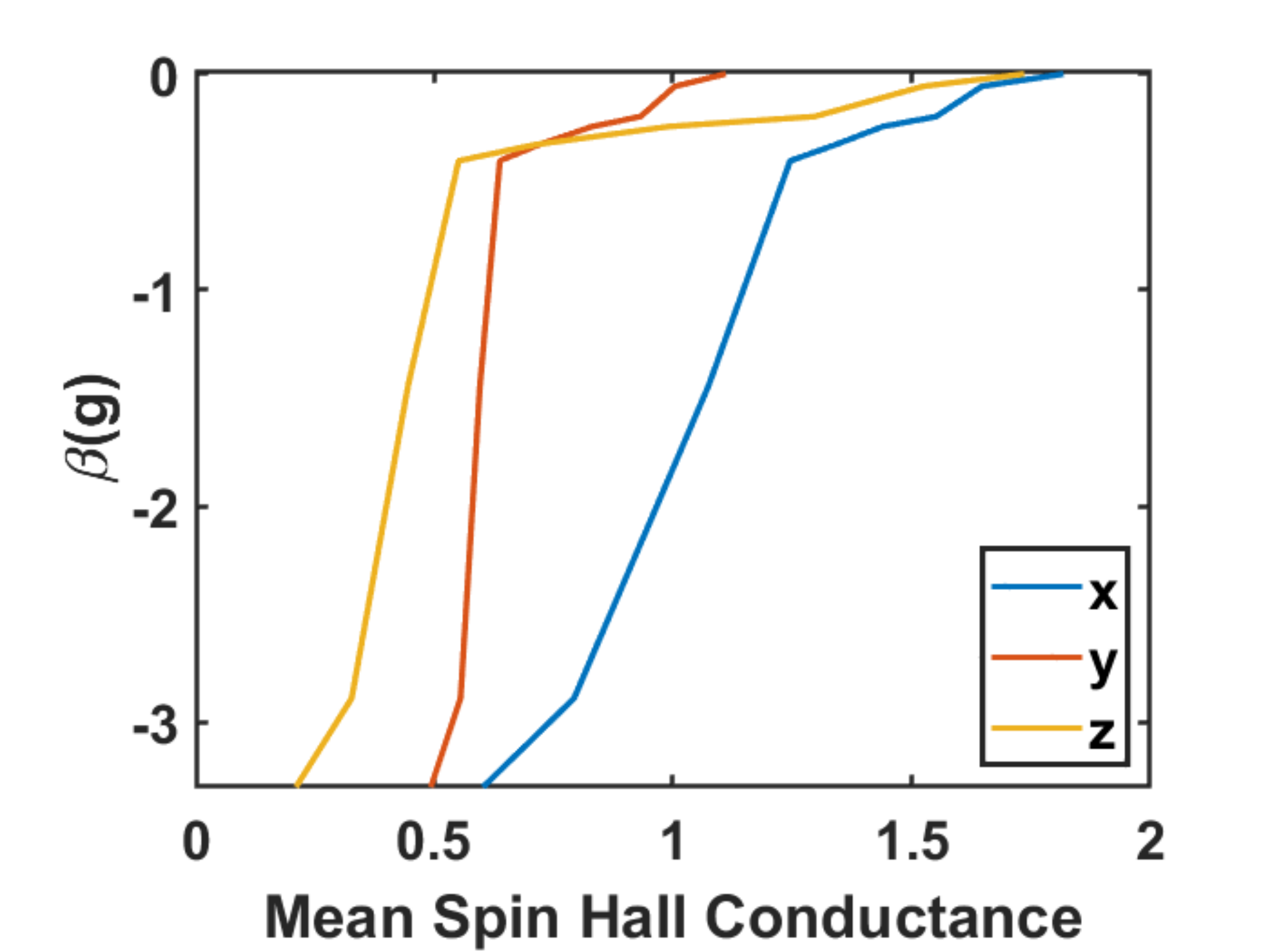}
     \caption{Plot of $\beta(g)$ as a function of mean spin Hall conductance for this $2D$ system along the polarization directions x, y and z. }
    \label{fig:P03_fig6}
  \end{minipage}\hfill\hspace{1.0cm}
 \end{figure}
\indent The mean and RMS spin Hall conductance along the z-direction is comparatively higher (within the range of $0.3-0.5$ in the measurable units) compared to the x- and y- directions, indicating preferential spin polarization along the out-of-plane direction in $WTe_2$. Results obtained for spin Hall conductance in this paper are consistent with the celebrated scaling theory of localization in a $2D$ system~\cite{abrahams1979scaling}. Motivated from this \enquote{gang of four} paper, we have shown the dependence between conductance (G) and length (L) of the device, by expressing:
\begin{equation}
    \beta(g)= \frac{dln(g)}{dln(L)}
\end{equation}
 where $g$ is defined as the generalized dimensionless conductance, termed as \enquote{Thouless Number}~\cite{abrahams1979scaling} and is expressed by: $g(L)= \frac{G(L)}{e^2/2\hbar}$. For $2D$ system, we infer from the fig.~\ref{fig:P03_fig6} that, for large values of $g$,  $\beta \approx 0$ for all the polarization angles, in fact $\beta$ is always negative. There is a universal crossover from logarithmic to exponential behavior. From the scaling theory prediction and from this pattern we extrapolate that, SHA reaches its maximum right where $\beta \approx 0$, marks the scale-invariant point estimated for the $2D$ system. Moreover, SHA can also be denoted by following equation~\cite{huang2013spin}:
\begin{equation}
\theta_{SHA}=\frac{2e}{\hbar}\frac{\sigma_{xy}^{SHC}}{\sigma_{xx}}
\end{equation}
here, $\sigma_{xy}^{SHC}$ is the spin Hall conductivity and $\sigma_{xx}$ is the longitudinal charge conductivity. In particular we observe that, the mean spin Hall conductance attains its maximum at intermediate disorder. This matches the predicted regime of strong quantum interference prior to Anderson localization, where $\sigma_{xx}$ is suppressed while $\sigma_{xy}^{SHC}$ remains robust. It may be noted that the manipulation of SHA via the ratio of between $\sigma_{xy}^{SHC}$ and $\sigma_{xx}$ has been discussed in a recent work of Das \textit{et al}~\cite{das2024origin} in connection with LSMMO/Pt bilayer, where the authors demonstrated that the spin-pumping effect can be suppressed by tuning the SHA. \\
\indent The strong anisotropy in hopping parameters plays a crucial role in determining the material's electronic band structure and topological properties. Overall, the variation pattern of current and SHA is consistent with already established spintronic phenomena and portrays the multifarious balance between disorder-driven localization and spin-dependent scattering processes in $WTe_2$ and TMDC-based designs.
\section{Conclusion} \label{section_4}
This paper is associated with the variation pattern of the mean and RMS values of charge current and spin current for varying disorder strength and energy, keeping uniform Rashba SOC values, and again for varying Rashba SOC values with a uniform energy value. These plots lead to the mean value and RMS fluctuations of SHA. Most of the variation patterns show a non-monotonic behavior with respect to an increase in disorder strength. We obtain a mean SHA of $25\%$ for energy variation and $30\%$ for Rashba SOC variation, which is well within the regime of the theoretical and experimental value of $WTe_2$  of different crystal structures. Also, we have shown the plots for mean and RMS spin Hall conductance along the polarized angles. The range of the mean conductance lies between 2 to 1.2 and that for the RMS spin Hall conductance lies between 0.35 to 0.22. More importantly, the results of this paper are in consistent with the celebrated scaling theory of localization in a $2D$ system, where the maximum SHA is obtained at the mobility edge. So, overall this comprehensive study is supported by rigorous numerical simulations using TB model and Kwant software. This study exploited the intrinsic spin characteristics of $WTe_2$ by controlling disorder strength, precise engineering of Rashba SOC and energy levels. It exhibits properties such as directional anisotropy, robustness, and stability, which are significant for potential spintronics applications and can be extended to extrinsic scattering mechanisms. 
\begin{acknowledgements}
The Research and Development work undertaken in the project under the Visvesvaraya Ph.D. Scheme of Ministry of Electronics and Information Technology, Government of India, is implemented by Digital India Corporation (formerly Media Lab Asia). This work was also supported by the Science and Engineering Research Board (SERB), Government of India,Grant No. CRG/2021/003102. The authors also wish to acknowledge the I-Hub Quantum Technology Foundation for the Chanakya Doctoral Fellowship (Grant No. I-HUB/DF/2021-22/002).
\end{acknowledgements}
\bibliography{reference}
\end{document}


\maketitle

\begin{figure*}[h]
\centering
\includegraphics[width=1.00\textwidth]{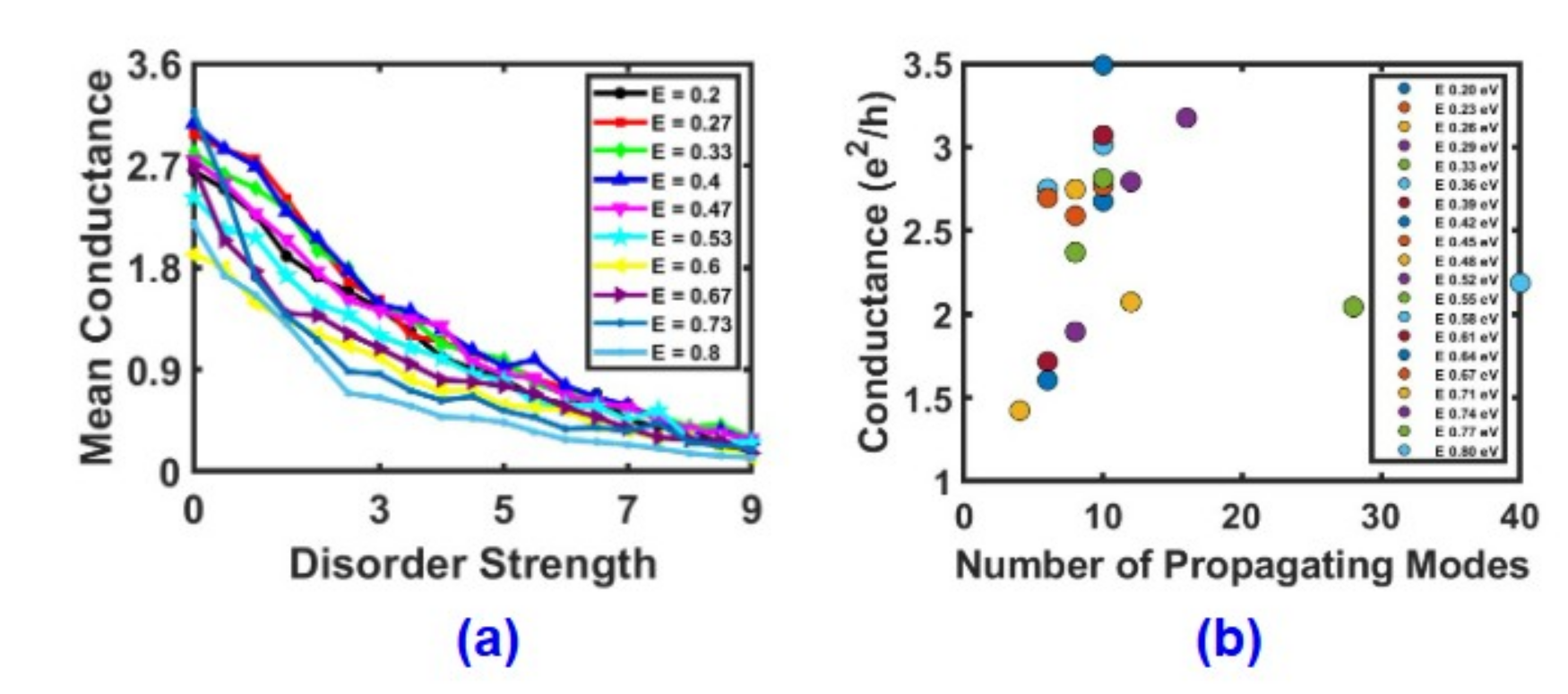}
\caption{(a) Plot of Mean conductance variation for different disorder strength and energy variation from 0.2 eV to 0.8 eV, (b) Plot of Conductance vs Number of propagating modes for energy variation from 0.2 eV to 0.8 eV.}
\label{fig:P03_figS1}
\end{figure*}

Fig.~\ref{fig:P03_figS1}(a) displays the mean conductance variation dependence on various disorder strength for energy variation from 0.2 eV to 0.8 eV. As we discussed in the manuscript, conductance pattern will follow the variation of mean charge current, the trend continued here. At low disorder, the conductance will have its maximum value and as the disorder starts increasing, conductance will gradually decrease and the plots converge at highest disorder strength. 

Fig.~\ref{fig:P03_figS1}(b) represents the dependence of conductance on the number of propagating modes for different energy values in a quantum transport system such as $WTe2$ devices. Number of propagating modes represent the number of available conducting channels (modes) and each mode acts as an independent pathway for electrons. More modes resulting in higher conductance. Conductance is measured in units of quantum conductance and it is standard in mesoscopic transport system. The scattered plot indicates, discrete increase in conductance with increasing number of modes. However, conductance values cluster around certain quantized values, reflecting the quantum nature of electron transport. As, the energy increases, the number of propagating modes generally increases and the individual points indicate specific configurations or states of the system.